\documentclass[12pt]{article}
\usepackage{epsfig}

\newcommand\pubnumber{SLAC-PUB-8855}
\newcommand\pubdate{\today}
\newcommand\hepnumber{hep-ph/0106046}

\def\csumb{Stanford Linear Accelerator Center\\
Stanford, California 94309, USA}
\def\support{\footnote{Work supported by
Department of Energy contract  DE--AC03--76SF00515.}} 

\def\Title#1{\begin{center} {\Large\bf #1 } \end{center}}
\def\Author#1{\begin{center}{ \sc #1} \end{center}}
\def\Address#1{\begin{center}{ \it #1} \end{center}}

\newcommand\pubblock{\rightline{\begin{tabular}{l} \pubnumber\\
         \pubdate\\ \hepnumber \end{tabular}}}
\newenvironment{Abstract}{\begin{quotation}  }{\end{quotation}}
\newenvironment{Presented}{\begin{quotation} \begin{center} 
             Presented at the\end{center}
      \begin{center}\begin{large}}{\end{large}\end{center} \end{quotation}}
\def\Acknowledgments{\bigskip  \bigskip \begin{center}
          \large\bf Acknowledgments\end{center}}

\makeatletter
\def\section{\@startsection{section}{0}{\z@}{5.5ex plus .5ex minus
 1.5ex}{2.3ex plus .2ex}{\large\bf}}
\def\subsection{\@startsection{subsection}{1}{\z@}{3.5ex plus .5ex minus
 1.5ex}{1.3ex plus .2ex}{\normalsize\bf}}
\def\subsubsection{\@startsection{subsubsection}{2}{\z@}{-3.5ex plus
-1ex minus  -.2ex}{2.3ex plus .2ex}{\normalsize\sl}}

\renewcommand{\@makecaption}[2]{%
   \vskip 10pt
   \setbox\@tempboxa\hbox{\small #1: #2}
   \ifdim \wd\@tempboxa >\hsize     % IF longer than one line:
       \small #1: #2\par          %   THEN set as ordinary paragraph.
     \else                        %   ELSE  center.
       \hbox to\hsize{\hfil\box\@tempboxa\hfil}
   \fi}

 \def\citenum#1{{\def\@cite##1##2{##1}\cite{#1}}}
\newcount\@tempcntc
\def\@citex[#1]#2{\if@filesw\immediate\write\@auxout{\string\citation{#2}}\fi
  \@tempcnta\z@\@tempcntb\m@ne\def\@citea{}\@cite{\@for\@citeb:=#2\do
    {\@ifundefined
       {b@\@citeb}{\@citeo\@tempcntb\m@ne\@citea\def\@citea{,}{\bf ?}\@warning
       {Citation `\@citeb' on page \thepage \space undefined}}%
    {\setbox\z@\hbox{\global\@tempcntc0\csname b@\@citeb\endcsname\relax}%
     \ifnum\@tempcntc=\z@ \@citeo\@tempcntb\m@ne
       \@citea\def\@citea{,}\hbox{\csname b@\@citeb\endcsname}%
     \else
      \advance\@tempcntb\@ne
      \ifnum\@tempcntb=\@tempcntc
      \else\advance\@tempcntb\m@ne\@citeo
      \@tempcnta\@tempcntc\@tempcntb\@tempcntc\fi\fi}}\@citeo}{#1}}
\def\@citeo{\ifnum\@tempcnta>\@tempcntb\else\@citea\def\@citea{,}%
  \ifnum\@tempcnta=\@tempcntb\the\@tempcnta\else
  {\advance\@tempcnta\@ne\ifnum\@tempcnta=\@tempcntb \else\def\@citea{--}\fi
    \advance\@tempcnta\m@ne\the\@tempcnta\@citea\the\@tempcntb}\fi\fi}
\makeatother

\def\beq{\begin{equation}}
\def\eeq#1{\label{#1}\end{equation}}
\def\eeqn{\end{equation}}

\newenvironment{Eqnarray}%
   {\arraycolsep 0.14em\begin{eqnarray}}{\end{eqnarray}}
\def\beqa{\begin{Eqnarray}}
\def\eeqa#1{\label{#1}\end{Eqnarray}}
\def\eeqan{\end{Eqnarray}}

\let\bar=\overbar

\def\Dslash{\not{\hbox{\kern-4pt $D$}}}
\def\dslash{\not{\hbox{\kern-2pt $\del$}}}

\def\ee{e^+e^-}

\def\msb{{\bar{\ssstyle M \kern -1pt S}}}

\def\lsim{\mathrel{\raise.3ex\hbox{$<$\kern-.75em\lower1ex\hbox{$\sim$}}}}
\def\gsim{\mathrel{\raise.3ex\hbox{$>$\kern-.75em\lower1ex\hbox{$\sim$}}}}

\newcommand{\evqq}     {\mbox{$ e \nu q \bar{q}$}}
\newcommand{\muvqq}     {\mbox{$ \mu \nu q \bar{q}$}}
\newcommand{\tauvqq}     {\mbox{$ \tau \nu q \bar{q}$}}
\newcommand{\qqqq}     {\mbox{$ q \bar{q} q \bar{q}$}}

\newcommand{\lvlv}     {\mbox{$ \ell \nu \ell \nu$}}
\def\dkg{\Delta\kappa_\gamma}
\def\ra{\rightarrow}
\def\ww{W^+W^-}
\def\wwl{W_{\rm L}W_{\rm L}}

\begin{document}
\begin{titlepage}
\pubblock

\vfill
\def\thefootnote{\fnsymbol{footnote}}
\Title{$\mathbf{WW}$ Physics at Future $\mathbf{\ee}$ Linear Colliders}
\vfill
\Author{Timothy L. Barklow\support}
\Address{\csumb}
\vfill
\begin{Abstract}
Measurements of triple gauge boson couplings and strong
electroweak symmetry breaking effects
at future $\ee$ linear colliders are  reviewed.   The results expected from a future $\ee$ linear collider are compared
with LHC expectations.
%with those from the LHC.     
%Complementary aspects of experimentation at the two colliders are noted.
\end{Abstract}
\vfill
\begin{Presented}
5th International Symposium on Radiative Corrections \\ 
(RADCOR--2000) \\[4pt]
Carmel CA, USA, 11--15 September, 2000
\end{Presented}
\vfill
\end{titlepage}
\def\thefootnote{\arabic{footnote}}
\setcounter{footnote}{0}

\section{Introduction}

%The process $\ee\ra\ww$ can provide insight
The measurement of gauge boson self-couplings at a future $\ee$ collider will provide insight
into new physics processes in the presence or absence
of new particle production.   In the absence of particle resonances, and in particular in the absence of 
a Higgs boson resonance, the measurement of gauge boson couplings 
will
provide a window to the new physics responsible for electroweak
symmetry breaking.   If there are many new particles being produced
-- if, for example, supersymmetric particles abound --
then the measurement of gauge boson couplings will prove valuable
since the gauge boson couplings will reflect 
the properties of the new particles through radiative corrections.

Experiments at LEP2 have demonstrated the viability of measuring gauge boson 
self-couplings at an $\ee$ collider.  Complex effects such as 
initial and final state radiation, ${\cal O}(\alpha)$ electroweak radiative corrections,
fragmentation, and detector bias are incorporated into analyses which utilize
all decay modes of the $W$ boson.  The present LEP2 triple gauge boson precision of 
a few percent~\cite{Jezequel:2000qr} exceeds the predictions for LEP2 sensitivity made a decade ago.

In this paper we review the prospect for studying triple gauge boson couplings and  strong
electroweak symmetry breaking effects at future $\ee$ linear colliders.   We will deal
primarily with the reaction $\ee\ra\ww$.  However, when discussing strong electroweak
symmetry breaking we will also consider the processes  $\ee\ra\nu \bar{\nu}\ww$, $\nu \bar{\nu}ZZ$, and $\nu \bar{\nu} t\bar{t}$.
Triple gauge boson production 
is important for the study of 
quartic gauge boson couplings, but is beyond the scope of this paper.

\section{Triple gauge boson couplings}

Gauge boson self-couplings include the triple gauge couplings (TGCs) 
and quartic gauge couplings (QGCs) 
of the photon, $W$ and $Z$.  Of special importance at a linear collider are 
the $WW\gamma$ and $WWZ$ TGCs
since a large sample of fully reconstructed $\ee\ra\ww$ events will be 
available to 
measure these couplings.           

   The effective Lagrangian for the general $\ww V$ vertex ($V=\gamma,Z$) 
contains 7 complex TGCs, denoted by $g^V_1$, $\kappa_V$, $\lambda_V$,
         $g^V_4$, $g^V_5$, $\tilde{\kappa}_V$, 
and $\tilde{\lambda}_V$~\cite{Hpzh:1987}.
The magnetic dipole and electric quadrupole moments of the 
$W$ are linear combinations of $\kappa_\gamma$ and $\lambda_\gamma$ while the
magnetic quadrupole and electric dipole moments are linear combinations of 
$\tilde{\kappa}_\gamma$ and $\tilde{\lambda}_\gamma$.  The TGCs $g^V_1$, 
$\kappa_V$, and $\lambda_V$
are C-- and P--conserving, $g^V_5$ is  C- and P-violating but conserves 
CP, and
$g^V_4$, $\tilde{\kappa}_V$, and $\tilde{\lambda}_V$ are CP-violating.
In the SM at tree--level all the TGCs are zero except
$g^V_1$=$\kappa_V$=1.  

If there is no Higgs boson resonance below about 800~GeV,  the 
interactions of the $W$ and $Z$ gauge bosons become strong above 1~TeV in the
$WW$, $WZ$ or $ZZ$ center-of-mass system.  In analogy 
with $\pi\pi$ scattering below the $\rho$ resonance,
the interactions of the $W$ and $Z$ bosons below the strong symmetry breaking resonances
can
be described by an effective chiral Lagrangian~\cite{Bagger:1993}.
These interactions induce
anomalous TGC's at tree-level:
\begin{eqnarray}
   \kappa_\gamma &=& 1+\frac{e^2}{32\pi^2s_w^2}\bigl(L_{9L}+L_{9R}\bigr) 
\nonumber
 \\  \kappa_Z  &=& 1+\frac{e^2}{32\pi^2s_w^2}
   \left(L_{9L}-\frac{s_w^2}{c_w^2}L_{9R}\right) \nonumber
 \\  g_1^Z  &=& 1+\frac{e^2}{32\pi^2s_w^2}\frac{L_{9L}}{c_w^2} \  \nonumber ,
\end{eqnarray}
where $s_w^2=\sin^2\theta_w$,
      $c_w^2=\cos^2\theta_w$, and  $L_{9L}$ and  $L_{9R}$
are chiral Lagrangian parameters.   
If we replace $L_{9L}$  and  $L_{9R}$ by the values of these parameters in
QCD, $\kappa_\gamma$  is shifted by  $\dkg \sim -3\times 10^{-3}$.

Standard Model radiative corrections~\cite{Ahn:1988fx} cause shifts in the 
TGCs of 
${\cal{O}}(10^{-4}-10^{-3})$ for CP--conserving
couplings and of ${\cal{O}}(10^{-10}-10^{-8})$ for CP-violating TGC's.  
Radiative corrections in the MSSM can cause shifts of 
${\cal{O}}(10^{-4}-10^{-2})$ 
in both the CP-conserving~\cite{Arhrib:1996dm} and CP-violating 
TGC's~\cite{Kitahara:1998bt}.

The methods used at LEP2 to measure TGCs provide a useful guide
to the measurement of TGCs at a linear collider.
When measuring TGCs
the kinematics of an $\ee\ra\ww$ event can be conveniently expressed in
 terms of 
the $\ww$ center-of-mass energy following initial state radiation (ISR), 
the masses of the $W^+$ and $W^-$, and five angles:  the angle 
between the  ${W^-}$ and initial
$e^-$ in the ${W^+W^-}$ rest frame, the polar and azimuthal
angles of the fermion in the rest frame of its parent $W^-$, and the polar
 and azimuthal
angles of the anti-fermion in the rest frame of its parent $W^+$.   

In practice not all of these
variables can be reconstructed unambiguously.   
For example, in events with hadronic decays it is often difficult to measure 
the flavor of the quark jet, and so there is usually a two-fold ambiguity 
for quark jet directions.
Also, it can be difficult to measure ISR and
consequently the measured $\ww$ center-of-mass energy is
often just the nominal $\sqrt{s}$. Monte Carlo simulation is used to account
 for 
detector resolution, quark hadronization, initial- and final-state radiation,
and other  effects.

The TGC measurement error at a linear collider can be estimated to a good
 approximation
by considering \evqq\ and \muvqq\ channels only, and by ignoring
all detector and radiation effects except for the requirement that 
the $\ww$ fiducial volume be restricted to  $|\cos{\theta_W}|<0.9$.    
Such an approach correctly predicts the TGC
sensitivity of LEP2 experiments and of detailed linear collider 
simulations~\cite{Burgard:1999}.
This rule-of-thumb approximation works because
LEP2 experiments and detailed linear collider simulations also use the 
\tauvqq\ , \lvlv\  and \qqqq\ channels, and the increased sensitivity 
from these 
extra channels makes up for the lost sensitivity due to
detector resolution, initial- and final-state radiation, and systematic errors.

\begin{table}[ht!]
\begin{center}
\begin{tabular}{|l|cc|cc|}\hline\hline
     & \multicolumn{4}{c|}{error $\times 10^{-4}$} \\
\hline
     & \multicolumn{2}{c|}{$\sqrt{s}=500$ GeV} &  
\multicolumn{2}{c|}{$\sqrt{s}=1000$ GeV} \\
 TGC & Re & Im & Re & Im \\
\hline
& & & & \\
$g^\gamma_1$ &  15.5    & 18.9     & 12.8     &  12.5     \\
$\kappa_\gamma$ &  \ 3.5    &  \ 9.8    & \ 1.2     &   \ 4.9    \\
$\lambda_\gamma$ & \ 5.4     &  \ 4.1    &  \ 2.0    &   \ 1.4    \\
$g^Z_1$          &  14.1    & 15.6     & 11.0     &  10.7     \\
$\kappa_Z$        & \ 3.8     &  \ 8.1    &  \ 1.4  &  \ 4.2    \\
$\lambda_Z$        &  \ 4.5     & \ 3.5     &  \ 1.7  &  \ 1.2  \\
\hline\hline
\end{tabular}
\caption{\label{tab:cp-conserving}
Expected errors for the real and imaginary parts of CP-conserving TGCs 
assuming $\sqrt{s}=500$~GeV, 
${\cal L}=500$~fb$^{-1}$ and  $\sqrt{s}=1000$~GeV, ${\cal L}=1000$~fb$^{-1}$.  
The results are for one-parameter fits in which all other 
TGCs are kept fixed at their SM values.}
\end{center}
%\end{table}
\vspace{.1in}
%\begin{table}[]
\begin{center}
\begin{tabular}{|l|cc|cc|}\hline\hline
     & \multicolumn{4}{c|}{error $\times 10^{-4}$} \\
\hline
     & \multicolumn{2}{c|}{$\sqrt{s}=500$ GeV} &  
\multicolumn{2}{c|}{$\sqrt{s}=1000$ GeV} \\
 TGC & Re & Im & Re & Im \\
\hline
& & & & \\
  $\tilde{\kappa}_\gamma$     & 22.5     &  16.4    &  14.9    &   12.0     \\
  $\tilde{\lambda}_\gamma$    & \ 5.8     & \ 4.0     & \ 2.0     & \ 1.4      \\
  $\tilde{\kappa}_Z$          & 17.3     &  13.8    & 11.8     &  10.3     \\
  $\tilde{\lambda}_Z$         & \ 4.6      & \ 3.4     & \ 1.7     & \ 1.2      \\
  $g^\gamma_4$                & 21.3     & 18.8     & 13.9     & 12.8      \\
  $g^\gamma_5$                & 19.3     & 21.6     & 13.3     & 13.4      \\
  $g^Z_4$                     & 17.9     & 15.2     & 12.0     & 10.4      \\
  $g^Z_5$                     & 16.0     & 16.7     & 11.4     & 10.7      \\
\hline\hline
\end{tabular}
\caption{\label{tab:cp-violating}
Expected errors for the real and imaginary parts of C- and P-violating 
TGCs assuming $\sqrt{s}=500$~GeV, 
${\cal L}=500$~fb$^{-1}$ and  $\sqrt{s}=1000$~GeV, ${\cal L}=1000$~fb$^{-1}$.  
The results are for
one-parameter fits in which all other TGCs are kept fixed at their SM values.}
\end{center}
\end{table}

Table~\ref{tab:cp-conserving} contains the estimates of the TGC
precision that can be obtained at $\sqrt{s}=500$ and 1000~GeV for the
CP-conserving couplings  $g^V_1$, $\kappa_V$, and $\lambda_V$.  
These estimates are derived
from one-parameter fits in which all other TGC parameters are kept fixed 
at their tree-level SM values.  
Table~\ref{tab:cp-violating} contains the corresponding estimates 
for the C- and P-violating couplings $\tilde{\kappa}_V$, 
$\tilde{\lambda}_V$, $g^V_4$, and $g^V_5$.  An alternative method of
measuring the $WW\gamma$ couplings is provided by the channel
$e^+e^-\rightarrow\nu\bar\nu\gamma$~\cite{k1}.

\begin{figure}[htp] % fig fig:gauge_lc_lhc
\centerline{\epsfig{file=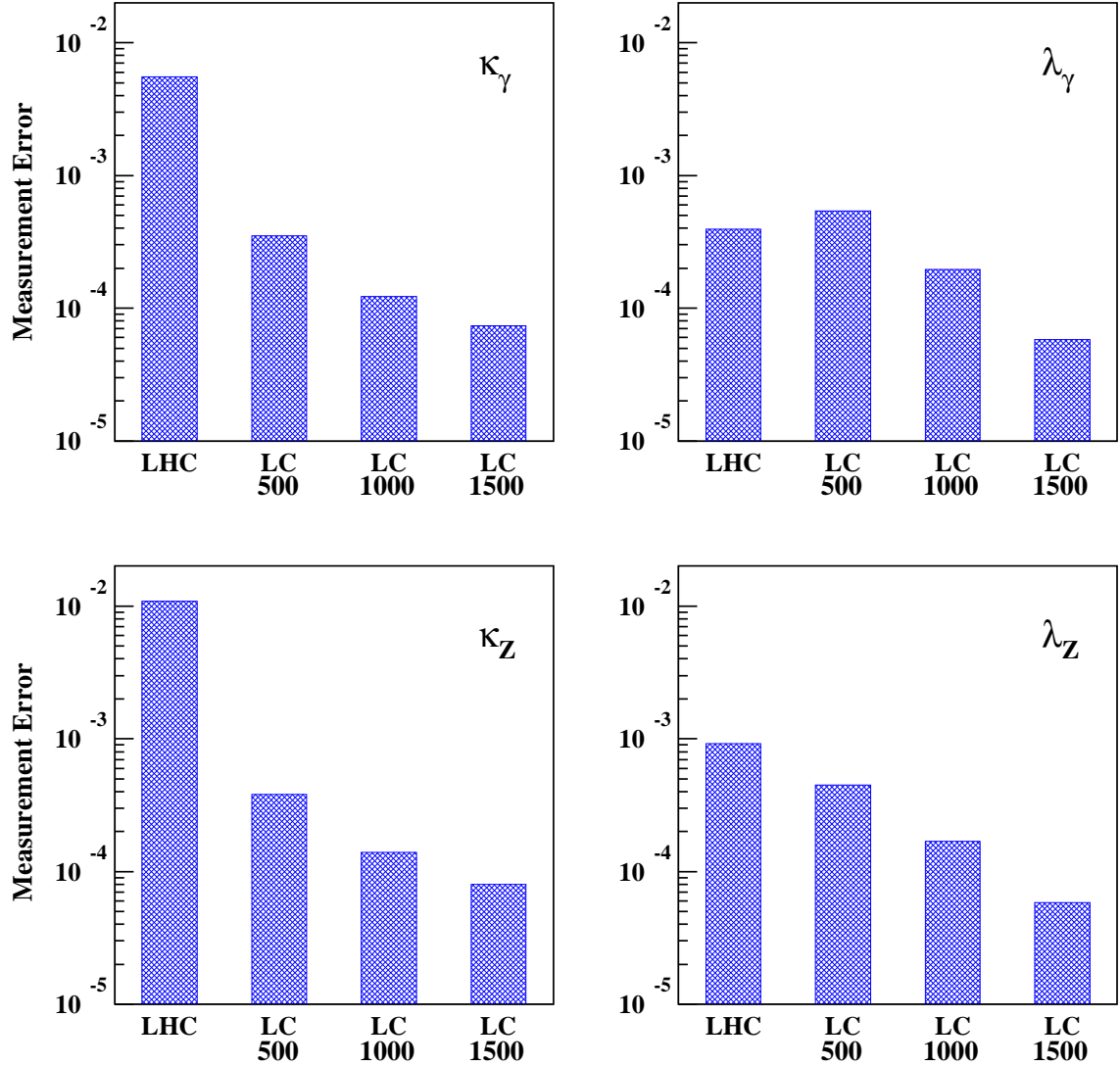,height=17cm}}
\vspace{10pt}
\caption{\label{fig:gauge_lc_lhc}
Expected measurement error for the real part of four different TGCs. 
The numbers below the ``LC'' labels refer to the
center-of-mass energy of the linear collider in GeV.
The luminosity of the LHC is assumed to be 300~$fb^{-1}$, while the 
luminosities
of the linear colliders are assumed to be  500, 1000, and 1000~$fb^{-1}$ for 
$\sqrt{s}$=500, 1000, and 1500~GeV respectively.}
\end{figure}

The difference in TGC precision between the LHC and a linear collider depends
 on the TGC, 
but typically the TGC precision at the linear collider will be substantially
 better,
even at $\sqrt{s}=500$~GeV.
Figure~\ref{fig:gauge_lc_lhc} shows the measurement
precision expected for the LHC~\cite{atlas:1999} and for linear colliders of
 three different
energies for four different TGCs.

\begin{figure}[htb] % fig fig:tgc_pair_corr
\centerline{\epsfig{file=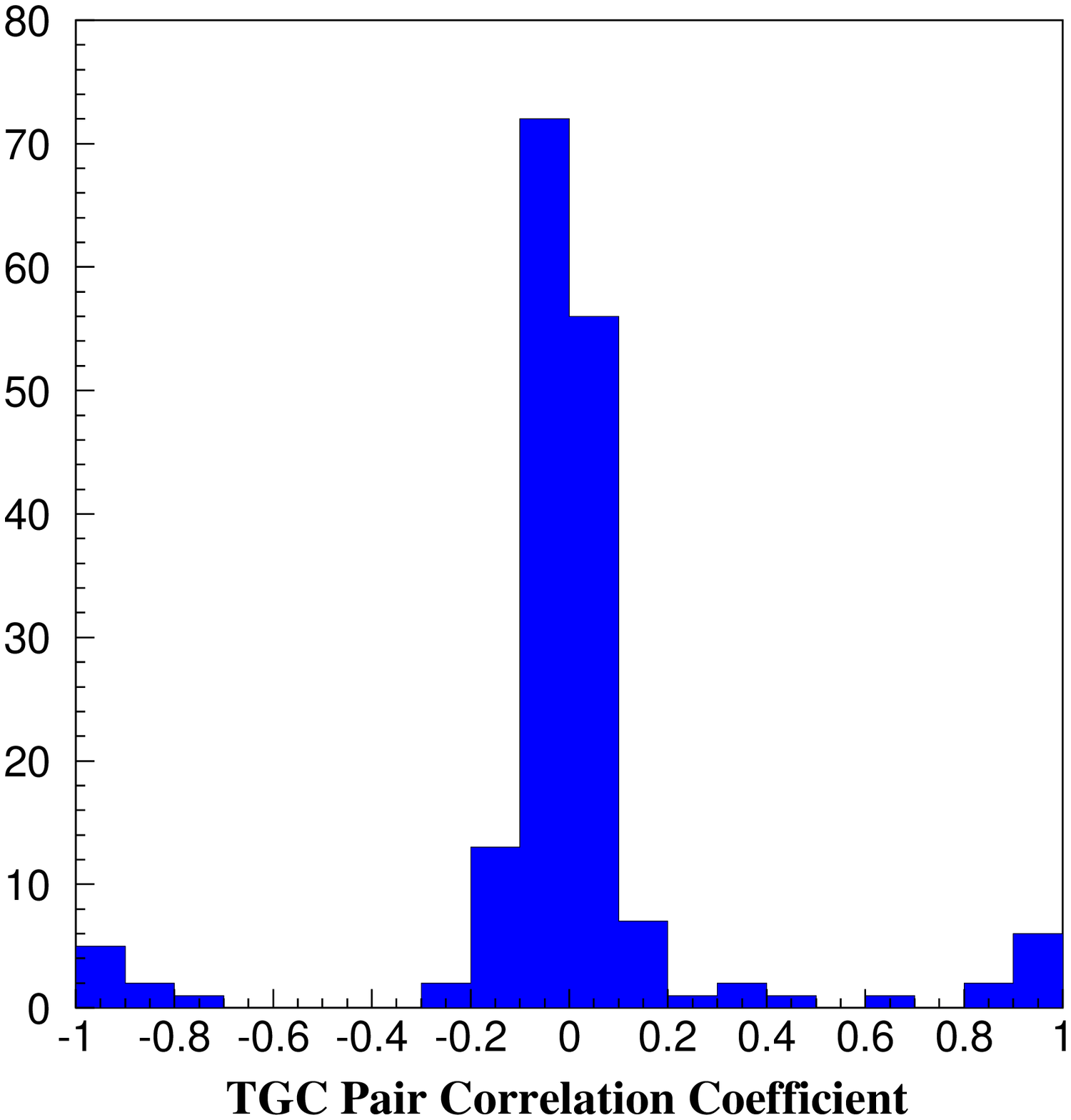,height=9cm}}
\vspace{10pt}
\caption{Histogram of correlation coefficients
for all 171 pairs of TGCs when 19 different TGCs are measured using
one-parameter fits at LEP2 (unpolarized beams).  The 19 TGCs are
made up of the real and imaginary
parts of the 8 C- and P-violating couplings along with the real parts of
the three CP-conserving couplings
$g^Z_1$, $\kappa_\gamma$,  $\lambda_\gamma$.}
\label{fig:tgc_pair_corr}
\end{figure}

If the goal of a TGC measurement program is to search for the  first sign of 
deviation from 
the SM, then one-parameter fits in which all other TGCs are kept fixed at their
 tree-level SM values
are certainly appropriate.  But what if the goal is to survey a large number
 TGCs, all of 
which seem to deviate from their SM value?  Is a 28-parameter fit required? 
 The answer is probably no,
as illustrated in Fig.~\ref{fig:tgc_pair_corr}.   

Figure~\ref{fig:tgc_pair_corr} shows the histogram of the correlation
 coefficients
for all 171 pairs of TGCs when 19 different TGCs are measured at LEP2 using 
one-parameter fits.  
The entries in Fig.~\ref{fig:tgc_pair_corr} with large positive correlations 
are pairs of TGCs 
that are related  to each 
other by the interchange
of $\gamma$ and $Z$.  The correlation between 
the two TGCs of each pair can be removed  using the dependence on electron beam polarization.
The entries in Fig.~\ref{fig:tgc_pair_corr} with large negative correlations 
are TGC pairs of the type 
$Re(\tilde{\kappa}_\gamma)/Re(\tilde{\lambda}_\gamma)$,
  $Re(\tilde{\kappa}_Z)/Re(\tilde{\lambda}_Z)$, etc.  Half of the TGC pairs 
with large negative
correlations will become uncorrelated once polarized electron beams are used,
  leaving only a small
number of TGC pairs with large negative or positive correlation coefficients.

\section{Strong $\mathbf{WW}$  scattering}

Strong $\ww$ scattering  can be studied at a linear collider with the
reactions $\ee\ra\nu \bar{\nu}\ww$, $\nu \bar{\nu}ZZ$, $\nu \bar{\nu} t\bar{t}$, and $\ww$~\cite{Barklow:1997nf}.
The final states $\nu \bar{\nu}\ww$, $\nu \bar{\nu}ZZ$ are used to study 
the I=J=0 channel in $\ww$ scattering, while the final state $\ww$ is best-suited
for studying the I=J=1 channel.   The $\nu \bar{\nu} t\bar{t}$ final state can be used to
investigate strong electroweak symmetry breaking in the fermion sector through the process
$\ww\ra t\bar{t}$.

The first step in studying strong $\ww$ scattering is
to separate the scattering of a pair of longitudinally polarized $W$'s, denoted by $\wwl$,
from transversely polarized $W$'s, and from background such as 
$\ee\ra\ee\ww$ and $e^- \bar{\nu}W^+Z$.
 Studies have shown that simple cuts can be used to achieve this separation 
in $\ee\ra\nu \bar{\nu}\ww$, $\nu \bar{\nu}ZZ$
at $\sqrt{s}=1000$~GeV, and that the
signals are comparable to those obtained at the LHC~\cite{Barger:1995cn}. 
Furthermore, by analyzing the gauge boson production
and decay angles it is possible to 
use these reactions to measure chiral Lagrangian parameters with an accuracy 
greater than that which can be achieved at the LHC~\cite{Chierici:2001}.

The reaction $\ee\ra \nu \bar{\nu} t\bar{t}$ provides unique access to 
$\ww\ra t\bar t$ since this process is overwhelmed 
by the background $gg\ra t\bar{t}$ at the LHC.  Techniques similar to those employed
to isolate $\wwl\ra \ww, ZZ$ can be used to measure
the enhancement in  $\wwl\ra t\bar{t}$ production~\cite{Barklow:1997wwtt}.
       Even in the absence
of a resonance it will be possible to clearly establish a
signal.  The ratio $S/\sqrt{B}$ is expected to be 12 for a linear
collider with $\sqrt{s} = 1$ TeV and 1000~fb$^{-1}$ and 80\%/0\%
electron/positron beam polarization,
increasing to
28 for the same data sample at $\sqrt{s} = 1.5$ TeV.

There are two approaches to studying strong $\ww$ scattering with
the process $\ee\ra \ww$.  
The first approach was discussed
in Section 2:  a strongly coupled gauge
boson sector induces anomalous TGCs which could be measured in 
$\ee\ra\ww$.  The precision of $4\times 10^{-4}$ 
 for the TGCs $\kappa_\gamma$ and $\kappa_Z$ at $\sqrt{s}=500$~GeV 
can be interpreted as a precision of $0.26$ for the chiral lagrangian parameters
$L_{9L}$ and $L_{9R}$.
Assuming naive dimensional analysis~\cite{Manohar:1984md}, 
such a measurement 
would provide a $8\sigma$ ($5\sigma$) signal for $L_{9L}$ and $L_{9R}$
if the strong symmetry breaking energy scale were 3~TeV (4~TeV).
The only drawback to this approach is that  the detection of anomalous TGCs does
not by itself provide unambiguous proof of strong electroweak symmetry breaking.   

The second approach involves an effect unique to 
strong $\ww$ scattering.
When $\ww$ scattering becomes strong
the amplitude for $\ee\ra\wwl$ develops a complex form factor $F_T$
in analogy with the pion form factor in $\ee\ra\pi^+\pi^-$~\cite{peskinomnes}.  
To evaluate the size of this effect the 
following expression for $F_T$ has been suggested:
\[
 F_T =
         \exp\left[{1\over \pi} \int_0^\infty
          ds'\delta(s',M_\rho,\Gamma_\rho)
          \left\{ {1\over s'-s-i\epsilon}-{1\over s'}\right\}
         \right]
\]
where
\[
\delta(s,M_\rho,\Gamma_\rho) = {1\over 96\pi} {s\over v^2}
+ {3\pi\over 8} \left[ \
\tanh (
{
s-M_\rho^2
\over
M_\rho\Gamma_\rho
}
)+1\right] \ .
\]
Here $M_\rho,\Gamma_\rho$ are the mass and width respectively of a vector 
resonance in $\wwl$ scattering.
The term 
\[
\delta(s) = {1\over 96\pi} {s\over v^2}
\]
is the Low Energy Theorem (LET) amplitude for $\wwl$ scattering at energies below a resonance.
Below the resonance
the real part of $F_T$ is proportional to $L_{9L}+L_{9R}$, and can therefore be interpreted as a TGC.   The imaginary part, however,
is a distinctive new effect.  

The real and imaginary parts of $F_T$ are measured~\cite{Barklow:2000let} in the 
same manner as the TGCs.
The $\ww$ production and decay angles are analyzed and an electron beam 
polarization of 80\% is assumed.
In contrast to TGCs, the analysis of $F_T$ seems to benefit from
even small amounts of jet flavor tagging. We therefore assume
that charm jets can be tagged with a
purity/efficiency of  100/33\%.
These purity/efficiency numbers are based on research~\cite{Damerell:1997} which indicates that
it may be possible to tag
charm jets with a 
purity/efficiency as high as 100/65\% given that
$b$ jet contamination is not a significant factor in $\ww$ pair-production and decay.

\begin{figure}[tbh] % fig fig:fteight
\centerline{\epsfig{file=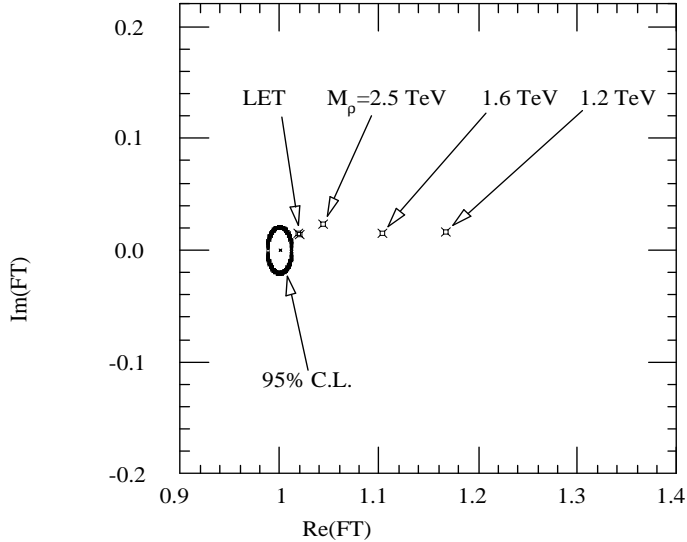,angle=-90,clip=,width=9cm}}
\vspace{10pt}
\caption{\label{fig:fteight}
95\% C.L. contour for $F_T$  for $\sqrt{s}=500$~GeV
and 500~$fb^{-1}$. Values of $F_T$  for various masses $M_\rho$ of a vector 
resonance in $\wwl$ scattering
are also shown. The $F_T$ point ``LET'' refers to the
case where no vector resonance exists at any mass in strong $\wwl$ scattering.}
\end{figure}

The expected 95\% confidence level limits for $F_T$ for $\sqrt{s}=500$~GeV
and a luminosity of 500~$fb^{-1}$ are shown in Fig. \ref{fig:fteight}, 
along with the predicted values of $F_T$ for various  masses $M_\rho$ of a vector 
resonance in $\wwl$ scattering.
The masses and widths of the vector resonances are chosen to coincide with 
those used in the ATLAS TDR~\cite{atlas:1999}.
The technipion form factor $F_T$ affects only the amplitude for $\ee\ra\wwl$, whereas TGCs
affect all amplitudes.   Through the use of electron beam polarization and the rich
angular information in $\ww$ production and decay, 
it will be possible to disentangle anomalous values of $F_T$ from 
other anomalous TGC values and deduce the mass of a strong vector resonance 
well below threshold, as suggested by Fig.~\ref{fig:fteight}.

\begin{figure}[] % fig fig:strong_lc_lhc
%\centerline{\epsfig{file=strong_lc_lhc_4x4.eps,height=17cm,width=16cm}}
\centerline{\epsfig{file=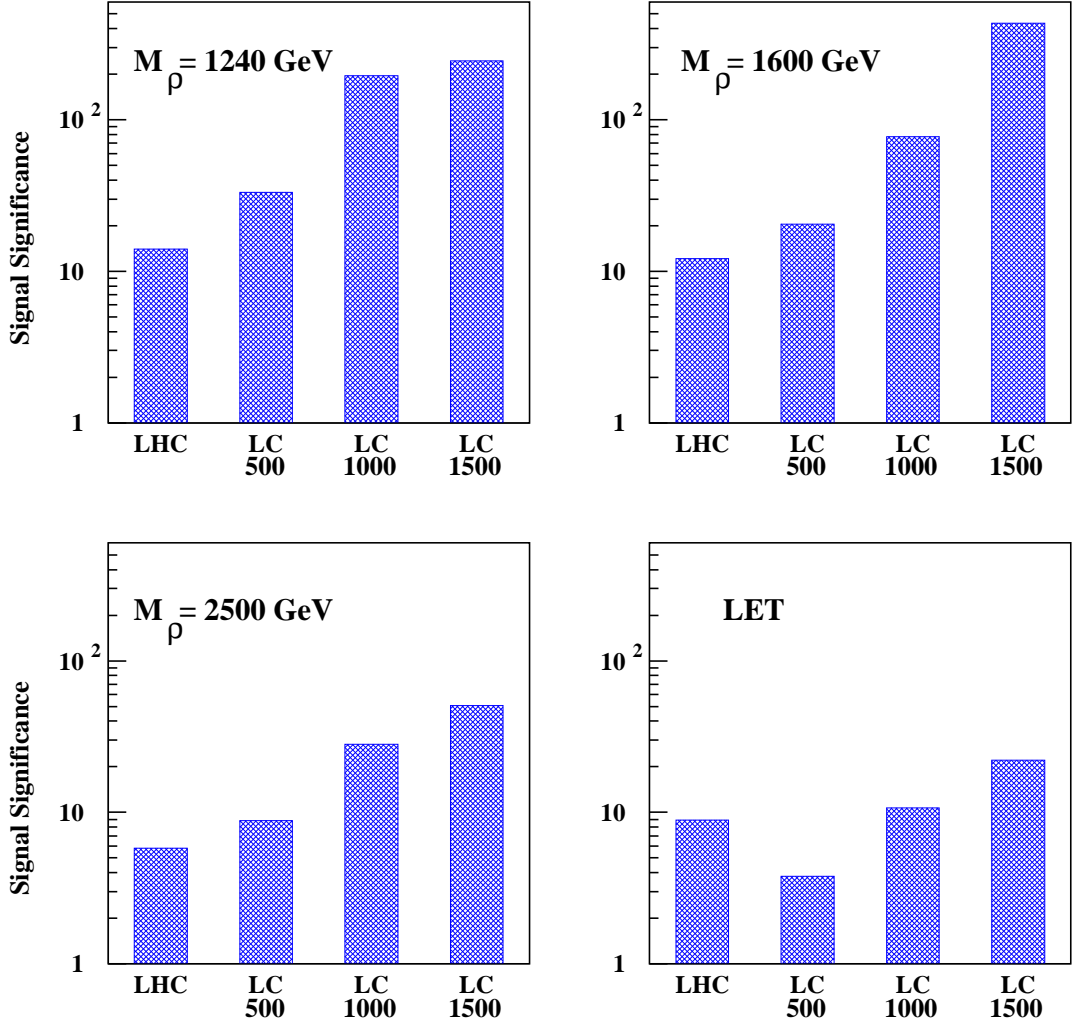,height=16cm,width=16cm}}
\vspace{10pt}
\caption{\label{fig:strong_lc_lhc}
Direct strong symmetry breaking signal significance in $\sigma$'s 
for various masses $M_\rho$ of a vector resonance in $\wwl$ scattering.
In the first three plots the signal at the LHC is a bump in the $WW$ cross
section; in the LET plot, the LHC signal is an enhancement over the SM
cross section.  The various LC signals are for enhancements of the
amplitude for pair-production of longitudinally polarized $W$ bosons.
The numbers below the ``LC'' labels refer to the
center-of-mass energy of the linear collider in GeV.
The luminosity of the LHC is assumed to be 300~$fb^{-1}$, while the luminosities
of the linear colliders are assumed to be  500, 1000, and 1000~$fb^{-1}$ for 
$\sqrt{s}$=500, 1000, and 1500~GeV respectively.  The lower right hand plot ``LET'' refers to the
case where no vector resonance exists at any mass in strong $\wwl$ scattering.}
\end{figure}

The signal significances obtained by combining the results for 
$\ee\ra\nu \bar{\nu}\ww$, $\nu \bar{\nu}ZZ$~\cite{Barger:1995cn} 
with the $F_T$ analysis of $\ww$~\cite{Barklow:2000let}
are displayed in Fig.~\ref{fig:strong_lc_lhc} along with the
results expected from the LHC~\cite{atlas:1999}.   
The LHC signal is a mass bump in $W^+W^-$; the LC signal is less
direct.  Nevertheless, the  signals at the LC are strong, particularly in $\ee\to W^+W^-$, where
the technirho effect gives a large enhancement of a very well-understood
Standard Model process.
Since the technipion 
form factor includes an integral
over the technirho resonance region, the linear collider signal 
significance is 
relatively insensitive to the technirho width.   (The real part of 
$F_T$ remains fixed 
as the width is varied, while the imaginary part grows as the width grows.)  
The LHC signal significance will drop
as the technirho width increases.  The large linear collider signals can 
be utilized to study a
vector resonance in detail; for example, 
the evolution of $F_T$ with $\hat{s}$ can be determined by measuring
the initial state radiation in $\ee\ra\ww$.

    Only when the vector resonance
disappears altogether (the LET case in the lower right-hand panel in Fig.~\ref{fig:strong_lc_lhc} ) 
does the  direct strong symmetry breaking signal from the $\sqrt{s}=500$~GeV linear collider
drop below the LHC signal.   At higher $\ee$ center-of-mass energies the linear collider 
signal exceeds the LHC signal.

\section{Conclusion}

A future $\ee$ linear collider operating in the center--of--mass energy
range of $0.5\ -\ 1.0$~TeV will measure TGCs with an accuracy of order
$10^{-4}$, which corresponds to an improvement of two orders of magnitude
over present LEP2 measurements and one order of magnitude over what
is expected from the LHC.   Such a precision is sufficient to test electroweak
radiative corrections to the TGCs.

Studies of strong electroweak symmetry breaking are enhanced by
a future $\ee$ linear collider.   
%The production mechanisms for $WW$ scattering,
%as well as the background, are limited to electroweak processes.
%As a consequence,  a measurement of a structureless enhancement in the total $WW$ scattering cross-section
%will have a smaller systematic error at an $\ee$ collider than at the LHC.
Signal and background in $WW$ scattering
are limited to electroweak processes, so that 
a measurement of a structureless enhancement in the total $WW$ scattering cross-section
will have a smaller systematic error at an $\ee$ collider than at the LHC.
%In addition, an $\ee$ collider can provide unique access to
%$\ww\ra t\bar{t}$, and can serve as a powerful probe of the $I=J=1$
%$WW$ scattering channel.
In addition, an $\ee$ collider does an excellent job measuring the lowest order 
parameters of the chiral lagrangian
for a strongly interacting gauge boson sector, as well as the
technipion
form factor for the pair-production of longitudinally polarized $W$ bosons.
Finally, an $\ee$ collider can provide unique access to the process $\ww\ra t\bar{t}$.

\Acknowledgments
I am grateful to Michael Peskin for many helpful discussions.


\begin{thebibliography}{99}

%%
%%  bibliographic items can be constructed using the LaTeX format in SPIRES:
%%    see    http://www.slac.stanford.edu/spires/hep/latex.html
%%  SPIRES will also supply the CITATION line information; please include it.
%%


%\cite{Jezequel:2000qr}
\bibitem{Jezequel:2000qr}
S.~Jezequel,
``Charged boson triple gauge couplings at LEP: W W gamma, W W Z and W  polarisation,''
LAPP-EXP-2000-10.


%\cite{Hpzh:1987}
\bibitem{Hpzh:1987} 
K.~Hagiwara, R.~D.~Peccei, D.~Zeppenfeld and K.~Hikasa,
Nucl.\ Phys.\  {\bf B282}, 253 (1987).



\bibitem{Bagger:1993} 
J. Bagger, S. Dawson, and G. Valencia,  Nucl.\ Phys.\ {\bf B399}, 264 (1993).

\bibitem{Ahn:1988fx}
C.~Ahn, M.~E.~Peskin, B.~W.~Lynn and S.~B.~Selipsky,
%``Delayed Unitarity Cancellation And Heavy Particle Effects In E+ E- $\to$ W+ W-,''
Nucl.\ Phys.  {\bf B309}, 221 (1988);
%%CITATION = NUPHA,B309,221;%%
% \bibitem{Fleischer:1992vq}
J.~Fleischer, J.~L.~Kneur, K.~Kolodziej, M.~Kuroda and D.~Schildknecht,
%``One loop improved Born approximation for e+ e- $\to$ W+ W-,''
Nucl.\ Phys.   {\bf B378}, 443 (1992)
[Erratum-ibid.  {\bf B426}, 443 (1992)].
%%CITATION = ERRAT,B426,246;%%


%\cite{Arhrib:1996dm}
\bibitem{Arhrib:1996dm}
A.~Arhrib, J.~L.~Kneur and G.~Moultaka,
%``MSSM radiative contributions to the WW$\gamma$ and WWZ form factors,''
Phys.\ Lett.  {\bf B376}, 127 (1996);
% [hep-ph/9512437].
%%CITATION = HEP-PH 9512437;%%
%\cite{Argyres:1996ib}
%\bibitem{Argyres:1996ib}
E.~N.~Argyres, A.~B.~Lahanas, C.~G.~Papadopoulos and V.~C.~Spanos,
%``Trilinear Gauge Boson Couplings in the MSSM,''
Phys.\ Lett.  {\bf B383}, 63 (1996). % [hep-ph/9603362];
%%CITATION = HEP-PH 9603362;%%
G. Couture, J.N. Ng, J.L. Hewett, and T.G. Rizzo, Phys.\ Rev.
{\bf D38}, 860 (1988).


%\cite{Kitahara:1998bt}
\bibitem{Kitahara:1998bt}
M.~Kitahara, M.~Marui, N.~Oshimo, T.~Saito and A.~Sugamoto,
%``CP-odd anomalous W boson couplings from supersymmetry,''
Eur.\ Phys.\ J.  {\bf C4}, 661 (1998), % [hep-ph/9710220].
%%CITATION = HEP-PH 9710220;%%

%\cite{Burgard:1999}
\bibitem{Burgard:1999}
C.~Burgard, in {\it Physics and Experiments with Future $\ee$ Linear Colliders}, 
%Sitges, Spain, 1999,
E. Fern\'{a}ndez and A. Pacheco, eds. (UAB Publications, Barcelona, 2000). 
%
%\bibitem{Menges:2001}
W.~Menges, ``A Study of Charge Current Triple Gauge Couplings at TESLA.'' LC-PHSM-2001-022. 
\verb+http://www.desy.de/~lcnotes+.

 \bibitem{k1}
D. Choudhury, J. Kalinowski, Nucl. Phys. {\bf B491}, 129 (1997).
D. Choudhury, J. Kalinowski, A. Kulesza, Phys. Lett. {\bf B457}, 193 (1999).

\bibitem{atlas:1999}
ATLAS Detector and Physics Performance Technical Design Report, LHCC 99-14/15 (1999).


%\cite{Barklow:1997nf}
\bibitem{Barklow:1997nf}
T.~L.~Barklow {\it et al.},
% ``Strong Coupling Electroweak Symmetry Breaking''
in {\it Proceedings of the 1996 DPF/DPB Summer Study On New Directions 
For High-Energy Physics} (Snowmass 96), 
D. G. Cassel, L. T. Gennari, and R. H. Siemann, eds. (SLAC,1997).
%``Strong coupling electroweak symmetry breaking,''
[hep-ph/9704217].
%%CITATION = HEP-PH 9704217;%%




%\cite{Barger:1995cn}
\bibitem{Barger:1995cn}
V.~Barger, K.~Cheung, T.~Han and R.~J.~Phillips,
%``Probing strongly interacting electroweak dynamics through W+ W- / Z Z 
% ratios at future e+ e- colliders,''
Phys.\ Rev.   {\bf D52}, 3815 (1995); % [hep-ph/9501379].
%%CITATION = HEP-PH 9501379;%%
%\cite{Boos:1998gw}
%\bibitem{Boos:1998gw}
E.~Boos, H.~J.~He, W.~Kilian, A.~Pukhov, C.~P.~Yuan and P.~M.~Zerwas,
%``Strongly interacting vector bosons at TeV e+- e- linear colliders,''
Phys.\ Rev.   {\bf D57}, 1553 (1998); % [hep-ph/9708310];
%%CITATION = HEP-PH 9708310;%%
%\cite{Boos:2000kj}
%E.~Boos, H.~J.~He, W.~Kilian, A.~Pukhov, C.~P.~Yuan and P.~M.~Zerwas,
%``Strongly interacting vector bosons at TeV e+- e- linear colliders.  (Addendum),''
Phys.\ Rev.   {\bf D61}, 077901 (2000). % [hep-ph/9908409].
%%CITATION = HEP-PH 9908409;%%

\bibitem{Chierici:2001}
R.~Chierici, S.~Rosati, M.~Kobel, ``Strong electroweak symmetry breaking 
signals in WW scattering at TESLA.'' LC-PHSM-2001-038 \verb+http://www.desy.de/~lcnotes+.


\bibitem{Barklow:1997wwtt}
T.~L.~Barklow, 
%``Using $\ee\ra\nu\bar \nu t\bar t$ to Probe Strong
%  Electroweak Symmetry Breaking at the NLC''
in {\it Proceedings of the 1996 DPF/DPB Summer Study On New Directions 
For High-Energy Physics} (Snowmass 96), 
D. G. Cassel, L. T. Gennari, and R. H. Siemann, eds. (SLAC,1997).
%
%\cite{RuizMorales:1999kz}
%\bibitem{RuizMorales:1999kz}
E.~Ruiz Morales and M.~E.~Peskin,
%``Probing strong electroweak symmetry breaking in W+ W- --> t anti-t,''
hep-ph/9909383.
%%CITATION = HEP-PH 9909383;%%
%
%\cite{Han:2001ic}
%\bibitem{Han:2001ic}
T.~Han, Y.~J.~Kim, A.~Likhoded and G.~Valencia,
%``Top-quark couplings to TeV resonances at future lepton colliders,''
Nucl.\ Phys.  {\bf B593}, 415 (2001); % [hep-ph/0005306].
%%CITATION = HEP-PH 0005306;%%
%
%\cite{Larios:2000xj}
%\bibitem{Larios:2000xj}
F.~Larios, T.~M.~Tait and C.~P.~Yuan,
%``The search for anomalous W+ W- t anti-t couplings at the LC,''
hep-ph/0101253.
%%CITATION = HEP-PH 0101253;%%]



%\cite{Manohar:1984md}
\bibitem{Manohar:1984md}
A.~Manohar and H.~Georgi,
%``Chiral Quarks And The Nonrelativistic Quark Model,''
Nucl.\ Phys.\ B {\bf 234}, 189 (1984).
%%CITATION = NUPHA,B234,189;%%
 
\bibitem{peskinomnes}
M. Peskin, in
{\it Physics in Collisions IV},
%Santa Cruz, CA, 1984,
A. Seiden, ed. (\'Editions Fronti\`eres, Gif-Sur-Yvette, France, 1984).
%
%\cite{Iddir:1990xn}
%\bibitem{Iddir:1990xn}
F.~Iddir, A.~Le Yaouanc, L.~Oliver, O.~Pene and J.~C.~Raynal,
%``W+ W- Production In E+ E- Colliders: A Test Of A Strongly Interacting Higgs Sector,''
Phys.\ Rev.   {\bf D41}, 22 (1990).
%%CITATION = PHRVA,D41,22;%% 

\bibitem{Barklow:2000let}
T.~L.~Barklow, ``LET Signals in $\ee\ra\ww$ at $\sqrt{s}=800$~GeV''
in Proceedings of 5th International Linear Collider Workshop (LCWS2000).

\bibitem{Damerell:1997}
C.~J.~S.~Damerell and D.~J.~Jackson, 
%``Vertex Detector Technology and 
%Jet Flavour Identification at the Future $\ee$ Linear Collider''
in {\it Proceedings of the 1996 DPF/DPB Summer Study On New Directions 
For High-Energy Physics} (Snowmass 96), 
D. G. Cassel, L. T. Gennari, and R. H. Siemann, eds. (SLAC,1997).





\end{thebibliography}
\end{document}